Astronomy
&
Astrophysics

# A multi-chord stellar occultation by the large trans-Neptunian object (174567) Varda[*]

D. Souami[1,2], F. Braga-Ribas[3,1,4,5], B. Sicardy[1], B. Morgado[1,5], J. L. Ortiz[6], J. Desmars[7,8],
J. I. B. Camargo[4,5], F. Vachier[8], J. Berthier[8], B. Carry[9], C. J. Anderson[10,11], R. Showers[11], K. Thomason[11],
P. D. Maley[10,12], W. Thomas[10], M. W. Buie[13], R. Leiva[13], J. M. Keller[14], R. Vieira-Martins[4,5,8], M. Assafin[15,5],
P. Santos-Sanz[6], N. Morales[6], R. Duffard[6], G. Benedetti-Rossi[1,5], A. R. Gomes-Júnior[16,5], R. Boufleur[4,5],
C. L. Pereira[3,5], G. Margoti[3], H. Pavlov[10,17], T. George[10], D. Oesper[10], J. Bardecker[10], R. Dunford[10,21], M. Kehrli[18],
C. Spencer[18], J. M. Cota[19], M. Garcia[19], C. Lara[19], K. A. McCandless[19], E. Self[19], J. Lecacheux[1], E. Frappa[20],
D. Dunham[10], and M. Emilio[22]

[1] LESIA UMR-8109, Observatoire de Paris, Université PSL, CNRS, Sorbonne Université, Univ. Paris Diderot, Sorbonne Paris Cité,
5 place Jules Janssen, 92195 Meudon, France
e-mail: damya.souami@obspm.fr
[2] naXys, University of Namur, Rempart de la Vierge, Namur 5000, Belgium
[3] Federal University of Technology – Paraná (UTFPR / DAFIS), Curitiba, Brazil
[4] Observatório Nacional/MCTIC, Rio de Janeiro, Brazil
[5] Laboratório Interinstitucional de e-Astronomia – LineA, Rua Gal. José Cristino 77, Rio de Janeiro, RJ 20921-400, Brazil
[6] Instituto de Astrofísica de Andalucía, IAA-CSIC, Glorieta de la Astronomía s/n, 18008 Granada, Spain
[7] Institut Polytechnique des Sciences Avancées IPSA, 63 boulevard de Brandebourg, 94200 Ivry-sur-Seine, France
[8] IMCCE-CNRS UMR8028, Observatoire de Paris, PSL Université, Sorbonne Université, 77 Av. Denfert-Rochereau, 75014 Paris,
France
[9] Université de la Côte d'Azur, Observatoire de la Côte d'Azur, CNRS, Laboratoire Lagrange, France
[10] International Occultation Timing Association (IOTA), PO Box 7152, WA 98042, USA
[11] College of Southern Idaho, Idaho, USA
[12] NASA Johnson Space Center Astronomical Society, Houston, TX, USA
[13] Southwest Research Institute, 1050 Walnut St., Suite 300, Boulder, CO 80302, USA
[14] University of Colorado, Boulder, Colorado, USA
[15] Observatório do Valongo/UFRJ, Rio de Janeiro, Brazil
[16] UNESP – São Paulo State University, Grupo de Dinâmica Orbital e Planetologia, Guaratinguetá, SP 12516-410, Brazil
[17] Tangra Observatory (E24), St. Clair, Australia
[18] California Polytechnic State University, San Luis Obispo, CA, USA
[19] Calipatria High School, Calipatria, CA, USA
[20] Euraster, 1 rue du tonnelier, 46100 Faycelles, France
[21] Jimginny Observatory (W08), Naperville, IL, USA
[22] Universidade Estadual de Ponta Grossa (UEPG), Ponta Grossa, Brazil

## ABSTRACT

*Context.* We present results from the first recorded stellar occultation by the large trans-Neptunian object (174567) Varda that was observed on September 10, 2018. Varda belongs to the high-inclination dynamically excited population, and has a satellite, Ilmarë, which is half the size of Varda.

*Aims.* We determine the size and albedo of Varda and constrain its 3D shape and density.

*Methods.* Thirteen different sites in the USA monitored the event, five of which detected an occultation by the main body. A best-fitting ellipse to the occultation chords provides the instantaneous limb of the body, from which the geometric albedo is computed. The size and shape of Varda are evaluated, and its bulk density is constrained using Varda's mass as is known from previous studies.

*Results.* The best-fitting elliptical limb has semi-major (equatorial) axis of $(383 \pm 3)$ km and an apparent oblateness of $0.066 \pm 0.047$, corresponding to an apparent area-equivalent radius $R'_{\text{equiv}} = (370 \pm 7)$ km and geometric albedo $p_v = 0.099 \pm 0.002$ assuming a visual absolute magnitude $H_V = 3.81 \pm 0.01$. Using three possible rotational periods for the body (4.76, 5.91, and 7.87 h), we derive corresponding MacLaurin solutions. Furthermore, given the low-amplitude ($0.06 \pm 0.01$) mag of the single-peaked rotational light-curve for the aforementioned periods, we consider the double periods. For the 5.91 h period (the most probable) and its double (11.82 h), we find bulk densities and true oblateness of $\rho = (1.78 \pm 0.06)$ g cm$^{-3}$, $\epsilon = 0.235 \pm 0.050$, and $\rho = (1.23 \pm 0.04)$ g cm$^{-3}$, $\epsilon = 0.080 \pm 0.049$. However, it must be noted that the other solutions cannot be excluded just yet.

**Key words.** methods: observational – occultations – Kuiper belt objects: individual: Varda

## 1. Introduction

Stellar occultations are one of the most accurate ground-based methods to directly determine the size and shape of Solar System objects, down to kilometric accuracies. In addition, these rare events can be used, amongst others, to reveal the existence of atmospheres, cometary activity, satellites, or ring systems around minor planets (Ortiz et al. 2020). We can cite here the discovery of rings around the Centaur (10199) Chariklo (Braga-Ribas et al. 2014) and the dwarf-planet (136108) Haumea (Ortiz et al. 2017) as well as the detection of the satellites of (87) Sylvia (Vachier et al. 2019).

In this paper, we report results obtained from a quintuple-chord occultation event involving trans-Neptunian object (174567) Varda (formerly 2003 MW$_{12}$). The event was observed on September 10, 2018, and is the first-ever observed stellar occultation by this body; it is used to constrain the size, 3D shape, geometric albedo, and bulk density of Varda. This campaign was carried out within the LUCKY STAR programme[1], which aims at studying and characterising objects of the outer Solar System, in particular, Centaurs and trans-Neptunian Objects (TNOs) in collaboration with the RECON project (Buie & Keller 2016).

Varda belongs to the hot classical TNO population. Moreover, it is a binary object; its satellite Ilmarë has about half its size (Grundy et al. 2015). The mass of the system can therefore be derived from Kepler's law and thus provides direct access to Varda's bulk density when the volume is determined using stellar occultations. Density measurements in turn contain clues for the conditions in the primordial solar system and/or the current internal structure of the body and other TNOs similar to Varda (Fernández 2020; Barucci & Merlin 2020).

A review of trans-Neptunian binaries (TNB) is given by Noll et al. (2020). These authors outline the current understanding and progress in the study of TNBs from the identification of binaries, determination of their mutual orbits, measurements of the system's mass, density, rotational state, component colours, mutual events, and the shapes. They reported 86 known TNBs at the time of writing. Today, 107 binary and multiple systems are known in the TNO population (including Pluto and Charon) that have one or more companions[2].

This paper is organised as follows. Section 2 provides an overview of the Varda-Ilmarë system. Section 3 describes our occultation prediction approach. Section 4 presents the data and the derived timings for each of the observed chords. In Sect. 5, we obtain the projected size and geometric albedo of Varda and constrain its 3D shape and density, and discuss the physical implications. Section 6 concludes the paper.

## 2. Overview of the Varda system

### 2.1. Physical properties of Varda

Varda was discovered on June 21, 2003, with the 0.9 m Spacewatch telescope at Kitt Peak Observatory (Larsen et al. 2007). A satellite was discovered around it in 2009, see Sect. 2.2. The physical properties of the Varda system available in the literature are summarised in Table 1. With an inclination of $i = 21.5°$ with respect to the ecliptic plane, Varda falls into the category of the dynamically hot population, which is characterised by inclinations larger than 5° (Fernández 2020).

Following the new TNO taxonomy, which is based on colour indices (B−V, V−R, V−I, V−J, V−H, and V−K; see Barucci et al. 2005; Fulchignoni et al. 2008), four classes are distinguished with increasingly red colours BB (neutral colour), BR, IR, and RR (very red). Using visible colour photometry and applying this new taxonomy, Perna et al. (2010) have re-classified Varda as an IR object. After conducting a study of 75 known Centaurs and TNOs, Barucci et al. (2011) found that IR-class objects belong to classical and resonant populations. None of the objects they identified in this class, including Varda, seemed to contain unambiguous water-ice signatures near 1.5 and 2 $\mu$m. Furthermore, Varda showed the largest positive slope in the 2.05–2.3 $\mu$m range. The spectrum obtained by Barucci et al. (2011) is consistent with the presence of ice tholin on Varda's surface.

### 2.2. Varda's satellite, Ilmarë

The *Hubble* Space Telescope (HST) Snapshot programme 11113 (conducted between July 2007 and April 2009) allowed the observation of 142 TNOs with the Wide Field and Planetary Camera 2 (WFPC2). HST images obtained on April 26, 2009, revealed the Varda satellite, Ilmarë, at a separation of about 0.12 arcsec (corresponding to ∼4000 km, i.e. about 12 Varda radii). Its size is about half that of Varda (Table 1). Additional (visible and near-IR) data acquired between April 2009 and July 2013 with the HST, the Keck II Telescope, and the Gemini North Telescope provided the relative offsets of Ilmarë with respect to Varda at twelve different epochs (Grundy et al. 2015 and references therein).

The derived orbit of Ilmarë has a mirror-ambiguity: two solutions are possible (cf. Table 1). The ambiguity should be removed by the end of 2020, when new relative positions will allow determining the correct solution. This will be possible due to the changing aspect of the orbit as seen from Earth.

Furthermore, Grundy et al. (2015) used their HST observations to perform separate photometry of the Varda-Ilmarë system (translated into B, V, I Johnson magnitudes). The B−V magnitudes show that the components are consistent with one another, whereas the V−I magnitudes show that Ilmarë is slightly redder than Varda (by (0.133 ± 0.062), see Grundy et al. 2015, and references therein). Moreover, Grundy et al. (2015) measured the difference in magnitude between the two components. It is $\Delta m = (1.734 \pm 0.042)$ mag.

### 2.3. Varda's rotational period

Thirouin et al. (2010, 2014) reported broadband CCD photometric observations carried out between 2006 and 2013 using several telescopes for the Varda-Ilmarë system. They give three possible solutions for Varda's rotational period (Table 1). Although their analysis of the low-amplitude (0.06 ± 0.01 mag) single-peaked light-curve favours the 5.91 h period, the study shows that two aliases exist at 4.76 and 7.87 h that cannot be discarded. All these periods are far shorter than the period of the mutual orbit, which is estimated at ∼138 h (5.75 days), see Grundy et al. (2015). These authors compiled all the photometric data from Thirouin et al. (2010, 2014) and re-evaluated the associated Lomb periodogramme (Lomb 1976; Press et al. 1992) in an attempt to identify frequencies that could/would be expected for tidally locked components in the system. The results were inconclusive.

## 3. Ephemerides and prediction of the occultation

Occultations by TNOs provide valuable opportunities of measuring their physical characteristics from ground-based

**Table 1.** Observational, physical, and orbital parameters of Varda.

| Quantity | Value | Ref. | Comments |
|---|---|---|---|
| Radius (km) | $353^{+41}_{-38}$ | V14 | Varda radius |
| | $396^{+46}_{-42}$ | V14 | Effective radius [(a)] from radiometric modelling (Herschel data) |
| | $361^{+41}_{-38}$ | G15 | Assuming that Varda and Ilmarë have spherical shapes and the same albedo, with brightness difference of $\Delta_{mag} = 1.734 \pm 0.042$, using photometric data of HST, Keck II, and Gemini North Telescopes, resulting in $R_{Varda} = 361^{+41}_{-38}$ km and $R_{Ilmarë} = 163^{+19}_{-17}$ km. |
| Mass ($10^{20}$ kg) | $2.664 \pm 0.064$ | G15 | System mass |
| Geometric albedo | $0.102^{+0.024}_{-0.020}$ | V14 | Radiometric modelling, Herschel data ("TNOs are Cool" programme) |
| Absolute magnitude ($H_V$) | $3.61 \pm 0.05$ | V14 | |
| | $3.988 \pm 0.048$ | A16 | with a phase coefficient [(b)] $\beta = -0.455 \pm 0.071$ |
| Rotational period (h) | 5.9/7.87 | T10 | Single-peaked light-curve with an amplitude of $0.06 \pm 0.01$ mag |
| | 5.91 | T14 | Other possible solutions (aliases) are mentioned: 4.76 h and 7.87 h |
| Spectral slope (%/100 nm) | $19.2 \pm 0.6$ | F09 | |
| Density $\rho$ (g cm$^{-3}$) | 1.12/0.63 | T10 | Associated with the periods (5.9 and 7.87 h) and sizes, respectively |
| Bulk density of the system | $1.27^{+0.41}_{-0.44}$ | V14 | Assuming that Varda and Ilmarë have equal albedo and density values |
| Varda-Ilmarë (g cm$^{-3}$) | $1.24^{+0.50}_{-0.35}$ | G15 | Using the mass and radius described above |
| *Ecliptic orbital elements of Varda* | | | |
| $a$ (AU) | 46.046 | LS | |
| $e$ | 0.140 | LS | NIMA ephemeris, see text |
| $i$ (deg) | 21.498 | LS | |

| | *Orbital elements of Ilmarë (ref:G15)* | |
|---|---|---|
| | solution 1 | solution 2 |
| $a$ (km) | $4812 \pm 35$ | $4805 \pm 35$ |
| $e$ | $0.0181 \pm 0.0045$ | $0.0247 \pm 0.0048$ |
| Orbital pole (J2000) | | |
| $\alpha_p$ (deg) | $273.0 \pm 1.5$ | $229.5 \pm 1.5$ |
| $\delta_p$ (deg) | $-11.0 \pm 1.9$ | $4.9 \pm 1.8$ |

**Notes.** [(a)] Vilenius et al. (2014) defined the radiometric (area-equivalent) effective diameter of a binary system as $D = \sqrt{D_1^2 + D_2^2}$, where $D_1$ and $D_2$ are the primary and secondary diameters, respectively. [(b)] This assumes a linear trend of the phase curves and considers a magnitude variability that is due to the rotational light-curve.
**References.** T10: Thirouin et al. (2010), T14: Thirouin et al. (2014), F09: Fornasier et al. (2009), V14: Vilenius et al. (2014), G15: Grundy et al. (2015), A16: Alvarez-Candal et al. (2016), LS : NIMA ephemerides (https://lesia.obspm.fr/lucky-star/obj.php?p=426).

observations. Huge efforts are dedicated to the predictions of events (Assafin et al. 2012; Camargo et al. 2014; Desmars et al. 2015). The second release of the *Gaia* stellar catalogue (Gaia Collaboration 2016, 2018) ensures an accuracy to the tenth of milliarcseconds (mas), which means that most of the uncertainty in the prediction resides in the ephemerides of the TNOs.

To obtain the most accurate ephemeris, regular astrometric observations are performed by our team in different observatories (ESO, Observatório do Pico dos Dias, Sierra Nevada, etc.). The astrometric reduction of these observations makes use of the *Gaia* catalogue, while orbits and ephemerides are obtained using the NIMA procedure (Desmars et al. 2015).

To predict this occultation, astrometric observations from MPC (1980-2018), ESO (2013), Observatório Pico dos Dias (2014, 2017, 2018), and Sierra Nevada (2018) were used to derive the NIMAv6 solution for Varda's orbit. As the position of Ilmarë around Varda was unknown, photo-centre and barycentre of the Varda system were assumed to be blended with Varda's centre in the NIMA orbit determination.

Figure 1 compares the NIMAv6 and JPL7 solutions in right ascension (weighted by $\cos \delta$) and in declination between 2013 and 2022. Residuals of observations from our survey are represented by mean points with their standard deviations as error bars (one point for observations during the same night). Dark blue points indicate data reduced with the WFI (Wide Field Imager) catalogue (Assafin et al. 2012; Camargo et al. 2014), light blue points are for data reduced with *Gaia*-DR1 and light red points are for data reduced with *Gaia*-DR2. NIMA also allows an estimation of the orbit uncertainty (1σ).

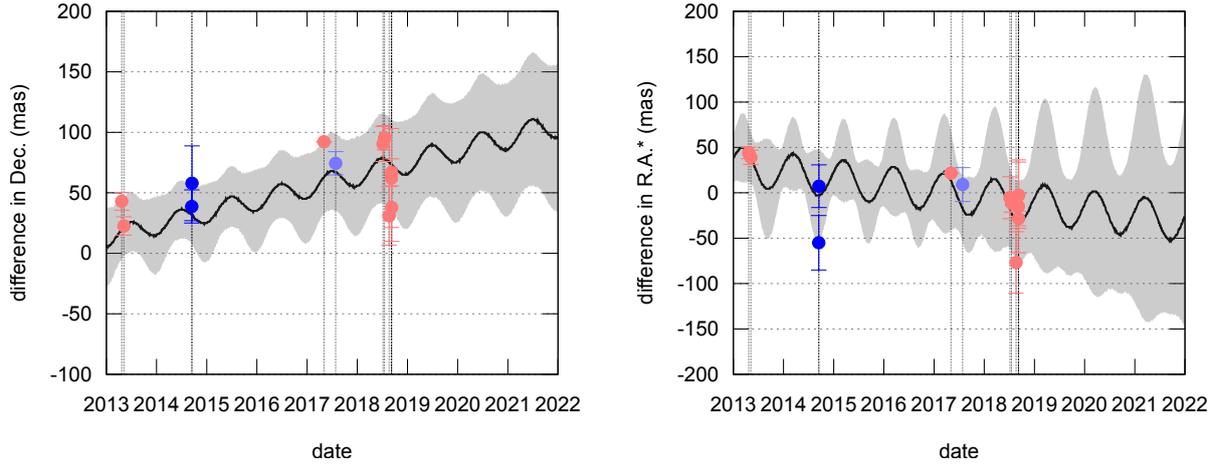

**Fig. 1.** Difference between NIMAv6 and JPL7 in Dec (declination, *left*) and RA* (right ascension weighted by cos δ, *right*) over the time interval 2013–2022. Each dot represents the mean residual and the standard deviation of one night of observations. Dark blue points indicate data reduced with the Wide Field Imager catalogue (see text), light blue points show data reduced with *Gaia*-DR1, and light red points show data reduced with *Gaia*-DR2. The continuous black line represents the best-fitting curve to these data, whereas the grey shaded area represents the 1σ uncertainty of the NIMAv6 ephemeris (https://lesia.obspm.fr/lucky-star/obj.php?p=46).

**Table 2.** Occultation circumstances.

| | |
|---|---|
| *Occultation prediction from NIMAv6* | |
| Date and time at geocentric close approach | September 10, 2018, 03:38:37 ± 117 s UT |
| Uncertainty on Varda position (RA and Dec) | $\Delta\alpha\cos(\delta) = 43.4$ mas, $\Delta\delta = 37.4$ mas |
| Along-track uncertainty | 1274 km (37.6 mas) |
| Across-track uncertainty | 1460 km (43.1 mas) |
| Geocentric shadow velocity | 10.86 km s⁻¹ |
| Maximum expected duration and magnitude drop | 66.7 s, 5.6 mag |
| Geocentric closest approach Varda-star | (85 ± 43) mas |
| *Occulted star (from Gaia-DR2)* | |
| Star source ID (stellar catalogue) | 4367203805493754752 |
| Geocentric star position (J2000) at the epoch | $\alpha = 17 : 18 : 25.12481 \pm 0.2$ mas, $\delta = -02 : 05 : 14.4282 \pm 0.1$ mas |
| G-mag/RP magnitude (mag) | 14.7/13.8 |
| Stellar diameter projected at the Varda distance | 0.82 km |
| *Varda at occultation epoch* | |
| Geocentric distance Δ | 46.6335 AU |
| Apparent diameter (physical diameter) | 21.4 mas (724 km) |
| Apparent bf V. magnitude | 20.3 |

When we use the NIMAv6 solution for the ephemeris and the *Gaia*-DR2 stellar position, the predicted geocentric mid-time for the occultation was on September 10, 2018, at 03:38:37 UT (±117 s). Table 2 provides the circumstances of the occultation, and Fig. 2 shows the prediction map.

This occultation was also predicted by RECON[3] using solely the Minor Planet Center data without additional astrometric positions. The predicted path[4], shifted to the west (~1500 km) compared to the Lucky Star prediction, highlights the importance of adding new astrometry and performing specific work to support the orbital fitting beyond using automatic tools. Our new astrometric positions for Varda are provided as supplementary material in the form of an ASCII file, and available in electronic form at the CDS. The final prediction was made available to observers through the Lucky Star webpage, the RECON webpage, and Occult Watcher[5].

## 4. Data analysis

Observations were performed at thirteen different stations in the USA (see the acronyms in Table 3). The observers were organised into eight RECON teams and five IOTA[6] (non-RECON) teams. The BDR team used a QHY174M-GPS, while all the other stations used the IOTA-VTI time stamp. All other relevant information regarding the observing circumstances is summarised in Table 3.

---

[3] Research and Education Collaborative Occultation Network (Buie & Keller 2016).

[4] https://www.boulder.swri.edu/~buie/recon/events/174567_180910_0154895.html

[5] https://www.occultwatcher.net/

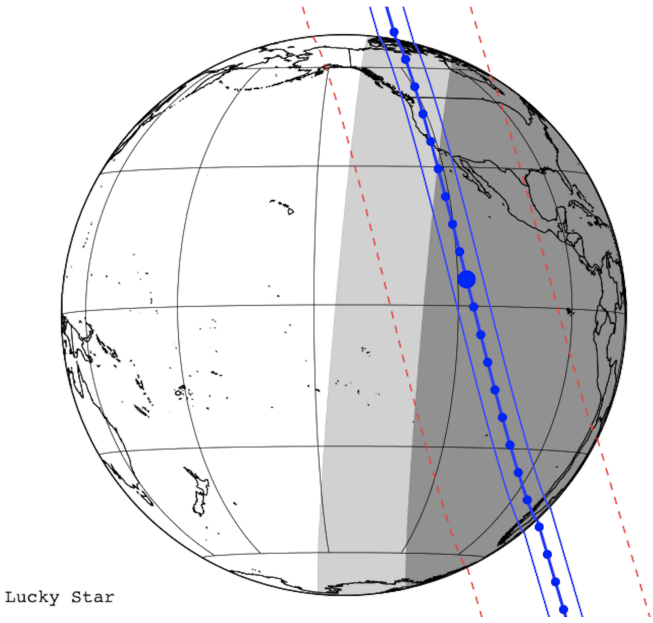

**Fig. 2.** Predicted shadow track for the September 10, 2018 stellar occultation by Varda; see the parameters in Table 2. Dark grey, light grey, and clear areas are in the night, twilight, and daytime, respectively. The Varda shadow is delimited by the continuous blue lines. It moves from north to south, and each bullet corresponds to one-minute intervals. The largest bullet corresponds to the geocentric closest approach. The red dashed lines show the $1\sigma$ path uncertainty, which is largely dominated by the uncertainties of the NIMAv6 ephemeris (Fig. 1) when compared to the position of the star (Table 2).

### 4.1. Occultation light-curves

Thirteen stations recorded data during the occultation, five of which (TWF, MHV, YMA, CAR, and FLO) reported positive detections. Differential aperture photometry was used to obtain the occultation light-curves, using the Platform for Reduction of Astronomical Images Automatically (PRAIA, Assafin et al. 2011). The light-curves from the five stations listed above are shown in Fig. 3; their photometric measurements are provided as supplementary material to this paper. Figure 3 presents these light-curves, to which we applied a manual offset in flux and centred at their respective mid-occultation times for the sake of legibility.

### 4.2. Ingress and egress timings

The start (ingress) and end (egress) times of the occultation were determined by fitting each event by a sharp opaque edge model,

---

[6] The International Occultation Timing Association has conducted a four-decade mission to observe and record eclipses of stars by minor planets and other bodies. Since the first announced observation, thousands of successful observations have been recorded. In the past decade, digital video records have been created with time sensitivities to 0.003 s. IOTA contributors did not receive funding and participated in this study on a voluntary basis.

after convolution by (1) Fresnel diffraction, (2) finite CCD bandwidth, (3) finite stellar diameter, and (4) finite integration time (see Braga-Ribas et al. 2013; Ortiz et al. 2017 for details). Fresnel diffraction operates over the scale $F = \sqrt{\lambda\Delta/2}$, where $\lambda$ is the observation wavelength and $\Delta$ is the geocentric distance of the object. From Table 2 we obtain $F \sim 0.5$ km in visible wavelengths. We estimate the stellar diameter to 0.8 km projected at the Varda distance, using $B$, $V$, $K$ magnitudes of 15.61, 14.32, 11.704 for the star (NOMAD catalogue, Zacharias et al. 2004), respectively, using the van Belle (1999) and Ochsenbein et al. (2000) formulae[7]. Given the geocentric shadow velocity of 10.86 km s$^{-1}$ and exposure times >1 s at all stations (see Tables 2 and 3), the main cause of light-curve smoothing is the finite integration time and not Fresnel diffraction or stellar diameter. However, the CAR station required further analysis, as an instrumental effect caused a smoothing of the light-curves over several data points (see Sect. 4.3), resulting in gradual ingress and egress, see discussion below. The occultation timings are obtained by minimising the classical $\chi^2$ function, using the same procedures as described in Sicardy et al. (2011). The best-fitting functions are displayed in Fig. 4, and the resulting chord details are listed in Table 4.

### 4.3. The CAR station

The CAR station light-curve (Fig. 5) shows a gradual ingress and egress of the star; the ingress and egress extend over more than 5 data points, i.e. over five seconds. Two scenarii were considered to explain this behaviour, (i) a Varda atmosphere, or (ii) an instrumental effect.

In case (i), a ray-tracing code was considered. It uses a nitrogen atmosphere with a temperature profile comparable to that of Pluto, with a rapid increase in temperature just above the surface that connects to an upper isothermal branch at ∼100 K. Then a surface pressure and temperature of ∼1 $\mu$bar and 38 K reproduce the profile of Fig. 5. Meanwhile, the signal-to-noise ratio (S/N) of the four other stations (TWF, MHV, YMA, and FLO) is insufficient and/or the time resolution is too low to test this model.

Case (ii): a closer examination of the CAR images revealed a slow motion of the stellar images. During ingress, the occulted star fades away, but its image no longer moves on the array, while the other sources continue to move clearly, indicating an instrumental effect (more details are given in Appendix B). To account for this effect, we added a non-instantaneous instrumental response in the modelling of the occultation light-curve. More precisely, we considered two possibilities for the response to an instantaneous light pulse:

a. In the first case, we assumed a simplified response that decays linearly over a time interval of $\Delta t_{resp}$. Exploring values of $\Delta t_{resp}$ between 0 and 10 s with steps of 0.01s, we obtained the occultation timings and the optimal $\Delta t_{resp}$ values timings by minimising the classical $\chi^2$ function, as we did for the other light-curves. We found the best (and satisfactory) simultaneous fit to the ingress and egress for $\Delta t_{resp} = 6.58^{+0.63}_{-0.52}$ s. (Fig. 5a).

b. In the second case, we assumed an (RC) electronic filter response ($e^{-t/\tau}$, where $\tau$ is the capacitor time constant) convolved with a rectangular function response that accounts for the finite integration time (cf. Table 3). We explored $\tau$ values in the interval [0,10] seconds with a step of 0.015 s. The occultation timings and the optimal $\tau$ values timings

---

**Table 3.** Observing circumstances of the 13 stations: 5 positive (+ve) chords, 6 negative (−ve), and 2 NDR (no data recorded) stations that experienced technical issues.

| Site | Topocentric coordinates | | Elevation | Telescope | Camera | Exposure [a] | Detection | Observers |
|------|-----------|----------|-----------|-----------|--------|--------------|-----------|-----------|
| | Longitude W | Latitude N | (m) | diameter (cm) | | time (s) | status | |
| CPSLO [*] | 120 39 36.0 | 35 18 01.8 | 119 | 28 | Mallincam 428 | 2.135 | −ve | M. Kehrli, C. Spencer, & M. Keidel |
| GAR [*] | 119 40 20.3 | 38 53 23.5 | 1502 | 30 | Watec 910HX | 1.068 | −ve | J. Bardecker |
| UMA [*] | 119 17 53.9 | 45 55 19.9 | 140 | 30 | Watec 910HX | 1.068 | −ve | T. George |
| IDY [*] | 116 42 42.0 | 33 44 03.0 | 1639 | 30.5 | Mallincam 428 | none | NDR | K. McArdle |
| CLP [*] | 115 31 28.0 | 33 07 30.0 | −56 | 28 | Mallincam 428 | none | NDR | [b] |
| MHV [*] | 114 35 48.9 | 35 01 54.1 | 184 | 28 | Mallincam 428 | 2.135 | +ve | J. White |
| TWF | 114 28 13.1 | 42 35 01.9 | 1133 | 60 | Watec 120N+ | 4.271 | +ve | C.J. Anderson |
| YMA [*] | 114 26 10.5 | 32 39 34.0 | 97 | 28 | Mallincam 428 | 2.135 | +ve | K. Conway, & D. Conway |
| CAR | 111 57 07.9 | 33 48 42.9 | 654 | 28 | Watec 910HX | 1.068 | +ve | P. D. Maley |
| FLO | 111 21 00.6 | 33 00 54.3 | 484 | 28 | Watec 120N+ | 1.068 | +ve | W. Thomas |
| BDR [*] | 105 09 46.8 | 40 15 09.0 | 1587 | 28 | QHY174M | 0.5 | −ve | M. Buie, J. Keller, & S. Haley |
| DOD | 090 08 31.1 | 42 57 36.9 | 390 | 30.5 | Watec 910HX | 1.068 | −ve | D. Oesper |
| NAP | 088 06 59.7 | 41 45 32.5 | 230 | 35 | QHY174M | 1 | −ve | B. Dunford |

**Notes.** [*] The sites annotated with an asterisk belong to the RECON network (Buie & Keller 2016). [a] The exposure time is equal to the cycle time. [b] CLP observers: A. McCandless, K. McCandless, X. Banaga, A. Carrillo, K. Hudson, D. Laguna, C. Lara, E. Self, S. Valdez, L. Torres, M. Garcia, J. Bustos, J. Cota. Acronyms: CPSLO: California Polytechnic State University, San Luis Obispo, California; GAR: Gardnerville, Nevada; UMA: Umatilla, Oregon; IDY: Idyllwild, California; CLP: Calipatria, California; MHV: Mohave Valley, Arizona; TWF: Twin Falls, Idaho; YMA: Yuma, Arizona; CAR: Carefree, Arizona; FLO: Florence, Arizona; BDR: Boulder, Colorado; DOD: Dodgeville, Wisconsin, and NAP: Naperville, Illinois.

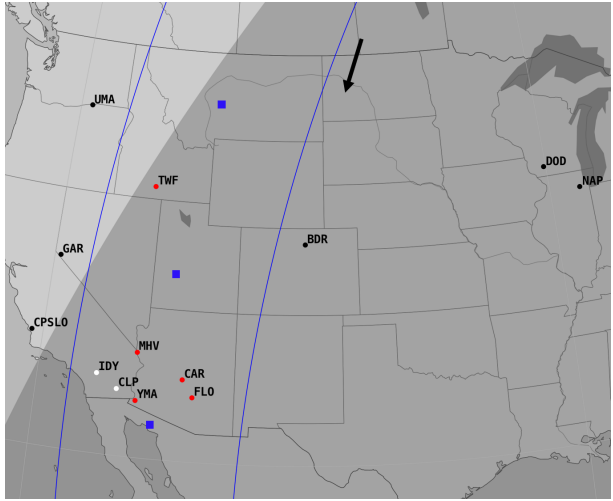

(a) Post-occultation map.

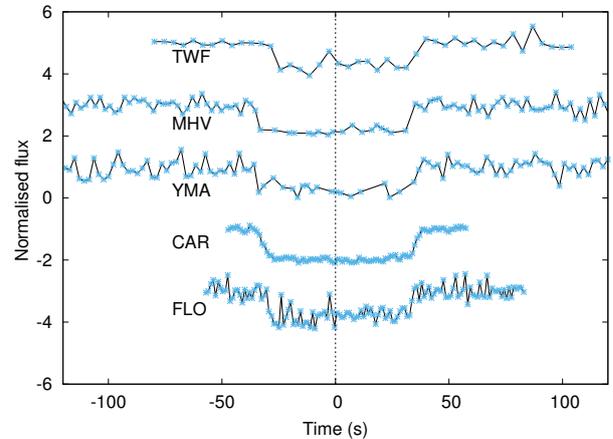

(b) Light-curves of the five positive chords.

**Fig. 3.** *Panel a*: post-occultation map for the stellar occultation by Varda on September 10, 2018. Blue lines represent the projected equivalent diameter of the object, and blue squares represent the position of the body centre (every minute); the northernmost dot corresponds to 03:33:02.934 UTC. The direction of the shadow is shown by the black arrow at the top centre of the panel. The red and black pins on the map represent the sites that reported +ve and −ve observations, whereas the white pins show the (NDR – no data recorded) stations that experienced technical problems that prevented them from recording the occultation (cf. Table 3). *Panel b*: five occultations detected at the TWF, MHV, YMA, CAR, and FLO stations (from the westernmost to the easternmost on the sky plane, cf. Fig. 6). Each light-curve has been normalised between 0 (Varda flux) and 1 (fluxes of the unocculted star and of Varda). For the sake of legibility, each light-curve has been centred at mid-chord time for each chord, and a manual offset in flux has been applied. The associated data to the normalised, non-centred light-curves are available at the CDS.

were obtained by minimising the classical $\chi^2$ function, as we did for the other light-curves. We found the best (and satisfactory) simultaneous fit to both ingress and egress for $\tau = 3.21 \pm 0.45$ s (Fig. 5b).

The resulting ingress and egress times obtained by the two methods agree at the $1\sigma$ confidence level. However, further analysis of the post-occultation data (cf. Appendix B) motivated our choice of an electrical RC filter (case b above). Finally, the overall uncertainty in the timing of this chord is mainly (∼60%) due to the uncertainty in the modelling of the (unknown) instrumental response. The gradual ingress and egress is clearly due to an instrumental effect.

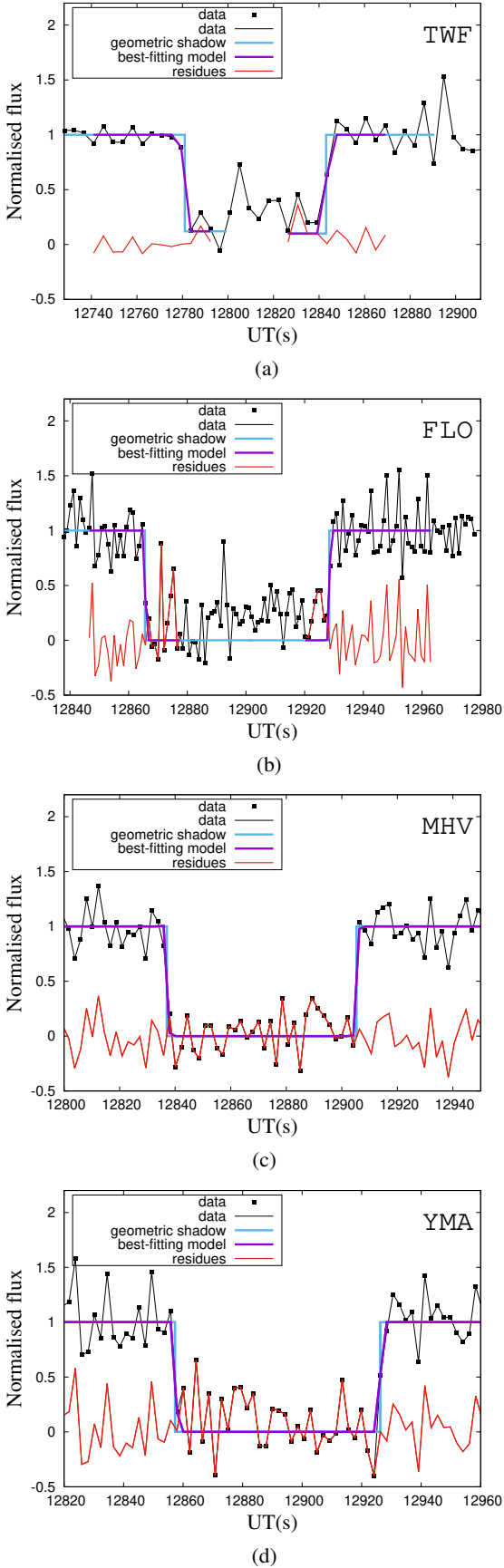

**Fig. 4.** Fit to ingress and egress at four stations. (*a*) Fit to ingress and egress for the TWF station. (*b*) Fit to ingress and egress for the FLO station. (*c*) Fit to ingress and egress for the MHV station. (*d*) Fit to ingress and egress for the YMA station.

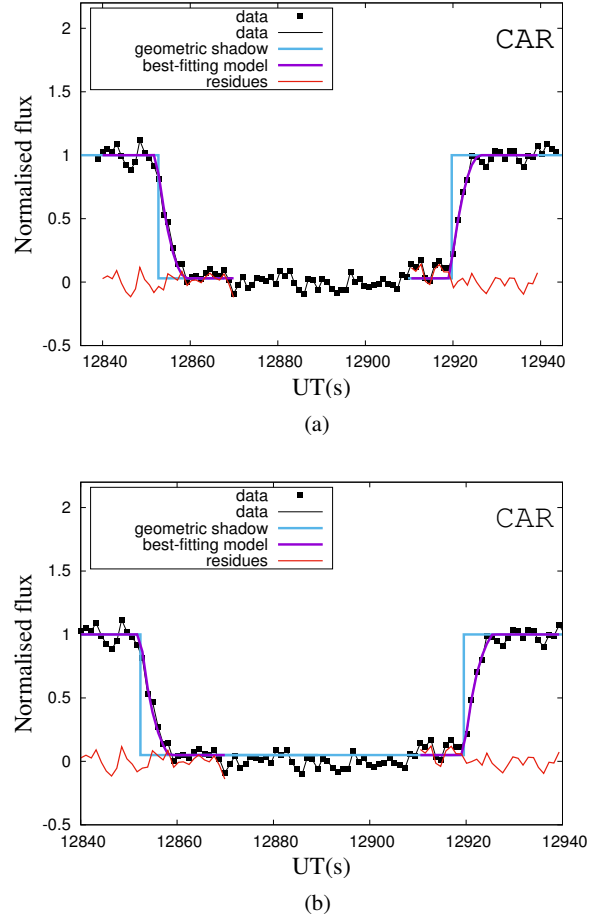

**Fig. 5.** Fitting of the CAR station chord. We note the gradual ingress and egress caused by the instrumental effect, see text for details. (*a*) Assuming a response that decays linearly over a time interval $\Delta\ t_{\text{resp}} = 6.58$ s (see text). (*b*) Assuming an (RC) electronic filter response with a time constant $\tau = 3.21$ s (see text).

The resulting egress and ingress timings as well as the duration and length of the chords at the five positive stations are listed in Table 4.

## 5. Results and discussions

### 5.1. Elliptical fit to the Varda limb

Because Varda's diameter is larger than 500 km (cf. Table 1), it is reasonable to assume that its limb is close to elliptical due to hydrostatic equilibrium (Tancredi & Favre 2008). Following the same procedures as in previous works (Braga-Ribas et al. 2013, 2014; Benedetti-Rossi et al. 2016, 2019; Dias-Oliveira et al. 2017; Ortiz et al. 2017), we fitted an ellipse to the chord extremities. The limb-fitting depends on five adjustable parameters:

i. the centre of the ellipse, $(f_c, g_c)$, which measures the offsets in $\alpha$ and $\delta$ that is to be applied (eastwards and northwards, respectively) to the adopted ephemeris (NIMAv6), assuming the stellar position of Table 2;

ii. the apparent semi-major axis $a'$;

iii. the apparent oblateness $\epsilon' = (a' - c')/a'$, where $c'$ is the apparent semi-minor axis;

**Table 4.** Signal-to-noise ratio per data point outside the occultation, and results of the fits displayed in Figs. 4 and 5 (timings, durations, and lengths of the chords).

| Site | Ingress time (UT hh:mm:ss.s) | Egress time (UT hh:mm:ss.s) | Chord duration (s) | Chord length (km) | $S/N$ per data point |
|------|------|------|------|------|------|
| TWF | 03:32:57.10 ± 0.60 | 03:33:58.83 ± 0.60 | 61.73 ± 0.85 | 670.4 ± 9.2 | 6.7 |
| MHV | 03:33:56.99 ± 0.76 | 03:35:05.27 ± 0.30 | 68.28 ± 0.82 | 741.5 ± 8.9 | 5.7 |
| YMA | 03:34:17.10 ± 0.60 | 03:35:26.39 ± 0.59 | 69.29 ± 0.84 | 752.5 ± 9.1 | 3.7 |
| CAR | 03:34:12.39 ± 0.47 | 03:35:19.52 ± 0.47 | 67.13 ± 0.66 | 729.0 ± 7.2 | 20 |
| FLO | 03:34:23.30 ± 0.13 | 03:35:26.75 ± 0.15 | 63.45 ± 0.20 | 689.1 ± 2.2 | 4.8 |

iv. the position angle $P'$ of the apparent semi-minor axis $c'$, measured eastwards from Celestial North.

These parameters were varied in wide ranges to determine the best-fitting limb. More precisely, we explored $-90° < P' < 90°$ in steps of $1°$, $0 < \epsilon' < 0.5$ in steps of 0.005, and $350\,\mathrm{km} < a' < 450\,\mathrm{km}$ in 1 km steps. For each value of $P'$, $\epsilon'$, and $a'$, we determined the best-fitting ellipse by adjusting the centre $(f_c, g_c)$. This best-fitting limb was obtained by minimising the function $\chi^2 = \sum_{i=1}^{N=6}(r_i - r_{\mathrm{ell},i})^2/\sigma_i^2$, where $r_i$ and $r_{\mathrm{ell},i}$ are the radial distance of the $i$th occultation point to the shadow centre and to the elliptical limb model, respectively. $\sigma_i$ is the radial uncertainty ($1\sigma$ level) stemming from the occultation timing uncertainties (Table 4). The $\chi^2$ value is based on the radial residual of the actual limb relative to the limb model. The associated error bar on each data point is then the timing uncertainty in the direction of the chord multiplied by the (radial) velocity of the star with respect to the centre of the body. It is therefore fully consistent with the approach that would minimise the residuals along the chords. Admittedly, this equivalence of methods works as long as the limb remains close to circular and none of the chords are close to grazing, which is the case here.

We define the value of $\chi^2$ per degree of freedom (or unbiased $\chi^2$) by $\chi^2_{\mathrm{pdf}} = \chi^2/(N - M)$, where $N = 10$ is the number of data points and $M = 5$ is the number of adjustable parameters. Figure 6 displays the results of these fits, and Table 5 provides the best-fitting adjusted parameters. The expected minimum value of $\chi^2_{\mathrm{pdf}}$ for a satisfactory fit is $\chi^2_{\mathrm{min_{pdf}}} = \frac{N-M}{M} = 1$, which is the case according to Table 5, noted $\chi^2_{\mathrm{pdf}} = 1.53$.

In our paper, the $1\sigma$ error bars on each fitted parameter stem from their respective marginal probabilities, and more precisely, their 68.3% confidence level values, independent of the other adjusted parameters. These domains are bound by the $\chi^2 < \chi^2_{\mathrm{min}} + 1$ criterion. This same approach was adopted in (Braga-Ribas et al. 2013, 2014; Benedetti-Rossi et al. 2016, 2019; Dias-Oliveira et al. 2017; Ortiz et al. 2017). The $\chi^2 < \chi^2_{\mathrm{min}} + 1$ limit corresponds to the confidence limit for one free parameter, while when all five free parameters are considered simultaneously, the formal $1\sigma$ confidence value for $\chi^2$ would write $\chi^2 < \chi^2_{\mathrm{min}} + 5.89$ (this is not the approach adopted here).

## 5.2. Size, shape, geometric albedo, and density of Varda

### 5.2.1. Geometric albedo $p_v$

We used the equivalent radius $R'_{\mathrm{equiv}}$ given in Table 5 and the absolute visual magnitude $H_V$ of Varda from the literature to derive the associated geometric albedo for Varda. The absolute magnitude given in the literature is that of the system Varda-Ilmarë (cf. Table 1). Moreover, as mentioned in Sect. 2.2, Grundy et al. (2015) have measured a difference in the visible magnitude between the two components of $\Delta m = (1.734 \pm 0.042)$ mag. It follows that the Varda absolute magnitude is $H_v = 3.81 \pm 0.01$. Using this latter value of the absolute magnitude, we derive the geometric albedo of Varda[8], $p_V = 0.099 \pm 0.002$. This value is consistent with but more accurate than the value from radiometric measurements, $p_V = 0.102^{+0.024}_{-0.020}$ reported by Vilenius et al. (2014). This shows that Varda is about as dark as the binary TNO 2003 $AZ_{84}$ (Dias-Oliveira et al. 2017).

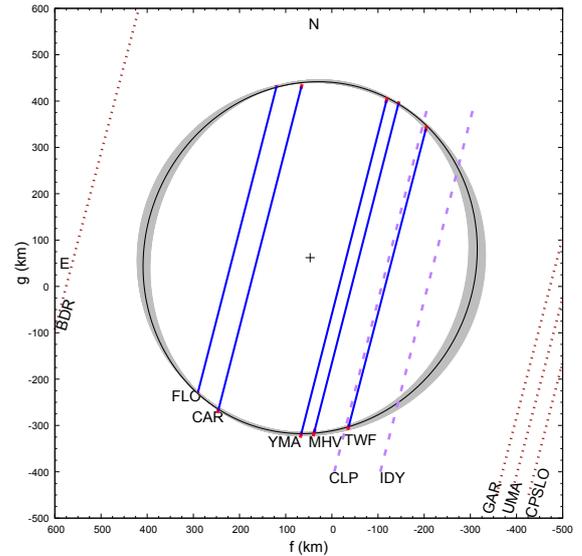

**Fig. 6.** Instantaneous best-fitting limb to the occultation chords. The blue lines are the occultation chords derived from Table 4, while the red segments are the uncertainties at the chord extremities. The black ellipse is the best-fitting solution, whose parameters are given in Table 5. The grey area shows all the solutions that are found to within a $1\sigma$ level from the best-fitting ellipse, i.e. corresponding to $\chi^2 < \chi^2_{\mathrm{min}} + 1$. The dashed purple lines show the projections onto the sky from the two NDR stations (cf. Table 3), and the brown dotted lines show the projections from the near-miss stations (cf. Fig. 3a).

Moreover, recent thermal emission data of the Varda-Ilmarë system observed with ALMA have shown that the ratio of the primary radius to the secondary radius is ∼2.10 (A. Moullet & E. Lellouch, 2020, priv. comm.). From the equivalent radius of Varda (cf. Table 5), we derive an equivalent radius of

---

[8] Using $p = (AU_{\mathrm{km}}/R_{\mathrm{equiv}})^2 10^{0.4(H_\odot - H)}$. Where $R_{\mathrm{equiv}}$ is expressed in kilometres, $H$ and $H_\odot = -26.74$ are the visual absolute magnitudes of Varda and the Sun, respectively, and $AU_{\mathrm{km}} = 149\,597\,870.7$ (Sicardy et al. 2011, supplementary information).

**Table 5.** Parameters of the elliptic fit to the limb.

| | |
|---|---|
| Apparent semi-major axis | $a' = (383 \pm 3)$ km |
| Apparent oblateness | $\epsilon' = 0.066 \pm 0.047$ |
| Equivalent radius [a] | $R'_{\text{equiv}} = (370 \pm 7)$ km |
| Geometric albedo [b] | $p_v = 0.099 \pm 0.002$ |
| Position angle | $P' = (67 \pm 8)$ deg |
| Best fit $\chi^2_{\text{pdf}}$ [c] | $\chi^2_{\text{pdf}} = 1.53$ |
| Varda J2000 geocentric position [d] | on September 10, 2018, at 03:34:00 UT |
| RA | 17 18 25.12972 ± 0.197 mas |
| Dec | −02 05 14.3055 ± 0.143 mas |

**Notes.** [a]The apparent area-equivalent radius is $R'_{\text{equiv}} = \sqrt{a'c'} = a'\sqrt{1-\epsilon'}$, where $c'$ is the apparent polar radius. [b]We do not know the rotational phase of Varda at the occultation epoch, therefore we cannot account for the peak-to-peak $\Delta\text{mag} = 0.06$ mag amplitude of the rotational light-curve (Thirouin et al. 2010). This adds an additional uncertainty of ±0.03 on $p$. [c]The calculation of $\chi^2_{\text{pdf}}$ is explained in the text. [d]The associated centre $(f_c, g_c) = (47.0 \pm 3.5, 62.0 \pm 1.6)$ km of the ellipse (Fig. 6) measures the offsets in $\alpha$ and $\delta$ that is to be applied to the adopted NIMAv6 ephemeris, which yields this new astrometric position.

~178 km for Ilmarë, which is about 15 km larger than the value reported by Grundy et al. (2015). Furthermore, using a difference $\Delta m = 1.734$ mag between the primary and the secondary implies that the albedo of Ilmarë is ~0.89 times that of Varda, that is, ~0.085.

### 5.2.2. Geometrical considerations

We now use the 2D limb fit obtained above to constrain the 3D shape of Varda. To do so, simplified assumptions must be made: Varda is an oblate, axisymmetric spheroid with an equatorial radius ($a$), polar radius ($c$), and true oblateness $\epsilon = (a-c)/a$. This choice is based on the small amplitude of the rotational light-curve ($\Delta\text{mag} = (0.06 \pm 0.01)$ mag[9], see Table 1), which favours a spheroid shape over a non-axisymmetric, triaxial ellipsoid (Thirouin et al. 2014), that is, the flux variations caused by albedo variegation.

The axisymmetric assumption leads to $a = a'$. Moreover, the apparent oblateness $\epsilon'$ is related to the true oblateness $\epsilon$ by the classical formula $c'^2 = a^2 \sin^2 B + c^2 \cos^2 B$ ($c'$ is the apparent polar radius),

$$(1-\epsilon')^2 = \sin^2 B + (1-\epsilon)^2 \cos^2 B, \qquad (1)$$

where $B$ is the planeto-centric sub-observer latitude, such that $B = 0°$ ($B = 90°$) corresponds to equator-on (pole-on) viewing geometry.

The density $\rho$ is then derived from the mass $M$ divided by the volume $3a^2c/4\pi$. From $c = a(1-\epsilon)$ and Eq. (1), we obtain

$$\rho = \frac{3M}{4\pi a'^3} \frac{\cos B}{\sqrt{(1-\epsilon')^2 - \sin^2 B}}. \qquad (2)$$

The mass $M$ is an observable quantity, while $a'$ and $\epsilon'$ are derived in the previous sub-section (see Table 5). Grundy et al. (2015) give the system mass (Varda plus Ilmarë) $M = (2.664 \pm 0.064) \times 10^{22}$ kg, where Varda accounts for 92% of the system mass. We adopt here a mass of $M = (2.45 \pm 0.06)10^{22}$ kg for Varda alone. As $B$ is a priori unknown, Eqs. (1) and (2) define a parametric curve $\epsilon(\rho)$ that is travelled as $B$ varies from 0 (the equator-on geometry with the minimum possible value of $\epsilon = \epsilon'$) to its maximum value $\arcsin(1 - \epsilon')$, corresponding to a completely (and unrealistic) flattened object, that is, $\epsilon = 1$. The function $\epsilon(\rho)$ is plotted in Fig. 7 as the dashed light blue line. The uncertainty in $\epsilon'$ causes a displacement of a given point $\rho(\epsilon)$ along this same curve, while the uncertainties in $M$ and $a'$ cause a transverse relative uncertainty in $\rho$ of

$$\left(\frac{\delta\rho}{\rho}\right)^2 = \left(\frac{\delta M}{M}\right)^2 + \left(3\frac{\delta a'}{a'}\right)^2, \qquad (3)$$

because $M$ and $a'$ are determined independently. The resulting error domain for $\rho$ is delimited by the light-blue continuous lines in Fig. 7. The minimum density provided by the spheroid model is $\rho \sim 1.19 \pm 0.04$ g cm$^{-3}$, up to an infinite value at $B = \arcsin(1 - \epsilon') \sim 69°$, corresponding to a flat object with $\epsilon = 1$.

### 5.2.3. MacLaurin solutions

We now proceed to assume that Varda is homogeneous and in hydrostatic equilibrium. This implies that Varda is either a MacLaurin spheroid or a Jacobi triaxial ellipsoid (Chandrasekhar 1987). With a typical equatorial radius of more than 380 km (Table 5) and its derived density (see below), Varda qualifies as a candidate for being a dwarf planet, i.e. a body in hydrostatic equilibrium (Tancredi & Favre 2008).

As mentioned earlier, an axisymmetric spheroid (here, a MacLaurin solution) is preferred over a triaxial ellipsoid (a Jacobi solution). In this case, the density $\rho$ of the body is related to its spin frequency $\omega = 2\pi/T_{\text{rot}}$ (where $T_{\text{rot}}$ is the rotational period) and its true oblateness $\epsilon$ by

$$\frac{\pi G\rho}{\omega^2} = \frac{\sin^2(\psi)\tan(\psi)}{2\psi[2 + \cos(2\psi)] - 3\sin(2\psi)}, \qquad (4)$$

where $\cos(\psi) = 1 - \epsilon$ and $G$ is the gravitational constant (Plummer 1919; Chandrasekhar 1987; Sicardy et al. 2011).

Three possible rotational periods (for single-peaked light-curves) are reported in the literature (see Sect. 2.3): 4.76 h, 5.91 h (the most probable, according to the authors), and 7.87 h. The resulting MacLaurin curves $\epsilon$ vs. $\rho$ are shown in Fig. 7 as purple, green, and red curves, respectively. The intersections of the light

---

[9] If we were to consider Varda (alone) with no satellite, the light-curve amplitude would slightly increase (as the difference in magnitude between the two components is $\Delta m = 1.734$) to ~0.07, and our assumptions would still be valid.

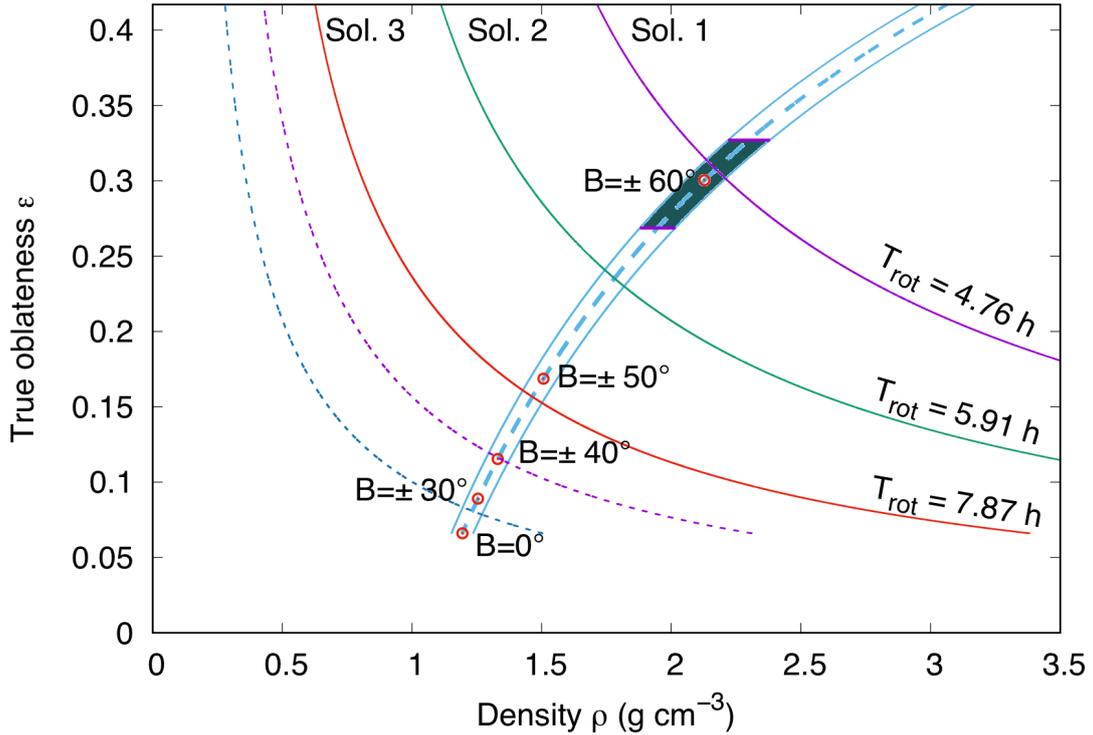

**Fig. 7.** Dashed light blue line: true oblateness ($\epsilon$) of Varda as a function of its density ($\rho$), as derived from our occultation results and knowledge of the mass of Varda (see Sect. 5.2.2 for details). The lowest point corresponds to the equator-on geometry ($B = 0$) for Varda, while the uppermost point (outside the figure) corresponds to $B = \arcsin(1 - \epsilon') \sim 69°$ for $\epsilon = 1$. The continuous blue lines to the left and right of the $\epsilon(\rho)$ curve delimit the $1\sigma$ uncertainty domain for the density. This domain stems from the uncertainties on $M$ and $a'$ (Eq. (3)). The MacLaurin solutions (Eq. (4)) corresponding to the rotational periods 4.76, 5.91, and 7.87 h (Thirouin et al. 2014) are shown as solid purple (solution 1), green (solution 2), and red (solution 3) curves, respectively, using the Varda mass from Sect. 5.2.2. The red circles along the curve $\epsilon(\rho)$ indicate some planeto-centric sub-observer latitudes $B$. The shaded region corresponds to one of the possible Ilmarë's orbit opening angle at epoch, $B_{open} = -59°8^{+1.4}_{-1.3}$. A solution compatible with an equatorial orbit for Ilmarë would be close to solution 2 and encompass solution 1 (shaded area). For each MacLaurin solution, we plot the corresponding solutions with the double period as dashed lines of the same colours, i.e. assuming double-peaked rotational light-curves (see Sect. 5.2.3 for details).

**Table 6.** MacLaurin solutions consistent with our occultation results.

| Rotation period (h) | $T = 4.76$ h (solution 1/1bis) | $T = 5.91$ h (solution 2/2bis) | $T = 7.87$ h (solution 3/3bis) |
|---|---|---|---|
| $\rho$ (g cm$^{-3}$) | $2.19 \pm 0.08/1.33 \pm 0.05$ | $1.78 \pm 0.06/1.23 \pm 0.04$ | $1.46 \pm 0.05$/no solution |
| $B$ (°) | $\sim 60/\sim 40$ | $\sim 56/\sim 25$ | $\sim 48$/no solution |
| $\epsilon$ | $0.307 \pm 0.074/0.116 \pm 0.047$ | $0.235 \pm 0.050/0.080 \pm 0.049$ | $0.157 \pm 0.047$/no solution |

**Notes.** This table gives the solutions for each of the three possible periods and their respective aliases (annotated with the suffix 'bis')

blue curve and the MacLaurin solutions in Fig. 7 provide solutions that are consistent with both our occultation results and the MacLaurin hypothesis.

All possible MacLaurin solutions consistent with our occultation's results are summarised in Table 6. Solutions 1, 2, and 3 are associated with the 4.71, 5.91, and 7.87 h rotational periods of the single-peaked light-curves, respectively; while solutions 1bis and 2bis are associated with the respective double periods 9.52 h, 11.81 h. There is no MacLaurin solution (3bis) associated with the double period 15.74 h. Although the light-curves associated with these double periods would mean that a body with a double-peaked rotational light-curve due to albedo features on opposite sides of the body, these solutions shall not be discarded because of the low-amplitude of the rotational light-curve (see previous sections). This stems from the extremely low-amplitude variability ($\Delta m < 0.15$ mag), for which it is almost impossible to distinguish the period from its double (Thirouin et al. 2014). Finally, we cannot exclude the existence of a non-axisymmetric, triaxial ellipsoid solution, in particular, a Jacobi-type body close to a pole-on orientation (to explain the low peak-to-peak amplitude of the light-curve of ~0.06 mag).

Except for solution 1 (for which $T = 4.76$ h), the bulk density values from all other solutions are consistent with but more accurate than the value $1.27^{+0.41}_{-0.44}$ g cm$^{-3}$ derived from thermal measurements for which equal albedo values and equal densities were assumed for both Varda and Ilmarë (Thirouin et al. 2014). Solution 1 gives a much higher and seemingly unlikely density for this size of object ($2.19 \pm 0.08$ g cm$^{-3}$), close to that of Quaoar. This solution, however, shall not excluded (see discussion in Sect. 5.3).

A further constraint can be considered by requiring an equatorial orbit for Ilmarë. This may, for example, be expected if Ilmarë resulted from the re-accretion of a collisional disc that surrounded Varda after an impact. The opening angle of its orbit, $B_{open}$, can be calculated from the orbital pole position (Table 1) and the astrometric position of Varda at this epoch.

We examine here the possibility that Ilmarë moves in the equatorial plane of one of the MacLaurin solutions considered above. Grundy et al. (2015) provided two (mirror-ambiguous) solutions for the Ilmarë orbital pole (Table 1). They yield two possible opening angles for the Ilmarë orbit as seen in the sky plane: $B_{open} = -74°.0^{+0.4}_{-0.2}$ and $B_{open} = -59°.8^{+1.4}_{-1.3}$. The first value is incompatible with all MacLaurin solutions considered here, see Fig. 7. Moreover, it would require an unrealistically high density for Varda, $\rho > 2.5$ g cm$^{-3}$. In contrast, the second Ilmare orbital solution is represented by the shaded area in Fig. 7. Moreover, it provides a position angle for the semi-minor axis of the orbit (as projected in the sky plane) of $96°.2^{+4.0}_{-3.6}$. This is consistent with the position angle $P' = (67° \pm 8°)$ of the Varda's limb semi-axis (Table 5), and thus, with an equatorial orbit for Ilmarë.

### 5.3. Discussion

The geometric albedo we derive for Varda ($p = 0.099 \pm 0.002$, see Table 5) can be compared to those of other TNOs. The dynamically hot population has a median geometric albedo of $0.085^{+0.084}_{-0.045}$ (Müller et al. 2020), excluding the Haumea family[10] and dwarf, planets noting that the median geometric albedo for cold classicals is $0.14^{+0.09}_{-0.07}$ (Ibid.). More recently, Müller et al. (2020) provided an overview of TNO physical properties at thermal wavelengths, in particular, for a sample of 26 hot classical TNOs (excluding Makemake, Haumea, and its family), with geometric albedo values ranging between 0.032 and 0.310, a median at 0.084, and a mean value at 0.102. Thus, the geometric albedo of Varda appears to be in line with those of other hot classical TNOs.

Our results are based on the assumptions that Varda is an oblate, axisymmetric spheroid with equatorial radius ($a$), polar radius ($c$), and true oblateness $\epsilon = (a - c)/a$. The small amplitude of the rotational single-peaked light-curve of Varda ($\Delta$mag $= 0.06 \pm 0.01$ mag, see Table 1) favours a spheroid solution over a non-axisymmetric triaxial ellipsoid, and motivates our search for MacLaurin solutions.

With an equatorial radius of about 380 km, Varda falls into the category of Kuiper belt objects (KBOs) that make a transition from small porous objects to dense KBOs, as defined in Bierson & Nimmo (2019). Simulations by these authors showed that this transition rapidly occurs for objects in the radius range 200–500 km. To explain the observed density distribution of the KBOs as a function of size, Bierson & Nimmo (2019) studied a sample of 11 of the 18 well-known TNBs using a 1D model that couples the KBOs thermal evolution with porosity evolution (where primordial porosity is removed over time). They concluded that the density distributions observed within the KBO populations are mainly a consequence of porosity, rather than mass. They also defined an input rock mass fraction parameter[11] $f_m = M_S/(M_S + M_i)$, where $M_S$ is the mass of silicates and $M_i$ is the mass of ice. In their model, Varda has $f_m > 70\%$, which is close to the value for Pluto and much lower than the values for Eris ($f_m \sim 90\%$) and Quaoar ($f_m \sim 85\%$), which are far larger and denser objects than Varda. This confirms that solution 1 ($\rho = 2.19 \pm 0.08$ g cm$^{-3}$) in Table 6 cannot be excluded just yet.

To summarise, solutions 1, 2, and 3 are compatible with a spheroid MacLaurin solution, but solution 1bis, and 2bis would be associated with triaxial ellipsoid shape of Varda. As discussed in Sect. 5.2.3, distinguishing a period from its double is nearly impossible for low-amplitude light-curves. As an example, we cite Ceres, whose oblate shape is well known from stellar occultations (i.e. 9.52 h, 11.82 h), and the DAWN spacecraft visit (Russell et al. 2016) while it exhibits a low-amplitude double-peaked rotational light-curve that is caused by albedo features (Chamberlain et al. 2007). In the trans-neptunian region, Makemake is also a clear example of an oblate body with a low-amplitude double-peaked light-curve (Hromakina et al. 2019).

If we were to assume that Varda is a body with a double-peaked light-curve caused by albedo features on opposite sides of the body, we should also consider the double of the aforementioned periods (i.e. 9.52 h, 11.82 h). The associated MacLaurin curves are plotted in Fig. 7 in purple, and blue, dashed lines, respectively. The associated density and oblateness values are given in Table 6.

### 6. Conclusion

We provided unique constraints on fundamental physical properties of Varda by refining its size, shape, geometric albedo, and bulk density. We used the first recorded stellar occultation by the TNO (174567) Varda that was observed from several stations in the US on September 10, 2018. This event resulted in five positive chords, from which an elliptical limb-fitting provides an apparent semi-major axis of $383 \pm 3$ km and an apparent oblateness of $0.066 \pm 0.047$ respectively, an area-equivalent radius of $370 \pm 7$ km, and a geometric albedo of $p_v = 0.099 \pm 0.002$.

We have derived five possible MacLaurin solutions for Varda, three assuming single-peaked rotational light-curves, and two assuming the associated respective double-peaked rotational light-curves. The associated density and oblateness values are listed in Table 6.

At this time, we cannot discriminate between these solutions. However, two solutions require our attention: the solution associated with the most probable period ($T = 5.91$ h) for a spheroid body with a single-peaked light-curve (solution 2: $\rho = 1.78 \pm 0.06$ g cm$^{-3}$ and $\epsilon = 0.235 \pm 0.050$), and its double period ($T = 11.82$ h) for a body with a double-peaked light-curve (solution 2bis: $\rho = 1.23 \pm 0.04$ g cm$^{-3}$ and $\epsilon = 0.080 \pm 0.049$). The relatively high densities derived for Varda (>1.5 g cm$^{-3}$) when we adopt the McLaurin solutions 1, 2, and 3 are consistent with low porosity for an object in this size range.

The gradual ingress and egress observed at one of the stations mimic the effect of an atmosphere, but are actually of an instrumental origin. Such effects could be observed again when similar set-ups are used, therefore some care is required to avoid erroneous interpretations.

From a more general standpoint, TNBs such as Varda are good occultation candidates to probe the outer Solar System because they provide accurate shapes (i.e. volumes) for bodies whose masses are well constrained from the satellite's motion. They provide a wide range of densities, from below that of water ice to that of nearly pure rock. These occultations constrain the formation scenarii for the Solar System and/or the current internal structure of these bodies.

---

[10] The Haumea family is characterised by high geometric albedo values, with a median $0.48^{+0.28}_{-0.18}$ (Müller et al. 2020).

[11] This is calibrated with a nominal value of 60% as this gives an object with $f_m = 70\%$ and a density of $\approx 750$ kg m$^{-3}$.

*Acknowledgment.* This campaign was carried out within the "Lucky Star" umbrella that agglomerates the efforts of the Paris, Granada and Rio teams. It is funded by the European Research Council under the European Community's H2020 (2014-2020/ERC Grant Agreement No. 669416). The following authors acknowledge the respective CNPq grants: F.B.-R. 309578/2017-5; R.V.-M. 304544/2017-5, 401903/2016-8; J.I.B.C. 308150/2016-3 and 305917/2019-6; M.A. 427700/2018-3, 310683/2017-3, 473002/2013-2. This study was financed in part by the Coordenação de Aperfeiçoamento de Pessoal de Nível Superior – Brasil (CAPES) – Finance Code 001 and the National Institute of Science and Technology of the e-Universe project (INCT do e-Universo, CNPq grant 465376/2014-2). G.B.R. acknowledges CAPES-FAPERJ/PAPDRJ grant E26/203.173/2016, MA FAPERJ grant E-26/111.488/2013 and ARGJr FAPESP grant 2018/11239-8. J.L.O., P.S.-S., N.M., and R.D. acknowledge financial support from the State Agency for Research of the Spanish MCIU through the "Center of Excellence Severo Ochoa" award for the Instituto de Astrofísica de Andalucía (SEV-2017-0709). P.S.-S. acknowledges financial support by the Spanish grant AYA-RTI2018-098657-J-I00 "LEO-SBNAF" (MCIU/AEI/FEDER, UE). Observations from the RECON network were provided by students, teachers, and community members, including Xavier Banaga, Jesus Bustos, Amanda Carrillo, Dorey W. Conway, Kenneth Conway, Danielle D. Laguna, Andrew E. McCandless, Kaitlin McArdle, and Jared T. White, Jr. The observers listed in this paper are but a small fraction of the total RECON network and their dedication to this project is deeply appreciated. Funding for RECON was provided by grants from NSF AST-1413287, AST-1413072, AST-1848621, and AST-1212159. This work has made use of data from the European Space Agency (ESA) mission *Gaia* (https://www.cosmos.esa.int/gaia), processed by the *Gaia* Data Processing and Analysis Consortium (DPAC, https://www.cosmos.esa.int/web/gaia/dpac/consortium). Funding for the DPAC has been provided by national institutions, in particular the institutions participating in the *Gaia* Multilateral Agreement. This research has made use of the VizieR catalogue access tool, CDS, Strasbourg, France (DOI: 10.26093/cds/vizier). The original description of the VizieR service was published in Ochsenbein et al. (2000).

# Appendix A: Astrometry

Astrometric observations of Varda were performed by our group using several telescopes between 2013 and 2018. The observations used the ESO 2.2 m Max Planck telescope at La Silla (IAU code 809), the 1.6 m telescope at Laboratório Nacional de Astronomia, Pico dos Dias Observatory (IAU code 874) as well as the 1.5 m telescope at Sierra Nevada Observatory (IAU code J86).

The astrometric positions following these observations are recorded in Table A.1. In this table, we indicate the date, the measured position (RA and Dec), the observation site (IAU code) as well as the astrometric catalogue that was used as reference for the reduction. Table A.1 is only available in electronic form at the CDS.

# Appendix B: Explaining the observed gradual ingress and egress in the CAR data

In Sect. 4.3 we modelled the observed gradual ingress and egress observed in the CAR station. The best-fitting model was obtained by introducing a non-instantaneous instrumental response. The latter is obtained by an (RC) electronic filter response ($e^{-t/\tau}$, where $\tau$ is the capacitor time constant) convolved with a rectangular function response that accounts for the finite integration time.

We attempt here to identify the physical or instrumental origin of this observed time delay. We first present an analysis of additional images taken a few minutes after the occultation, and conclude by giving a reason that explains the observed phenomenon.

Analysis of the after-event video segment ($\sim$03:43 UT, about 5 min after the occultation) showed a gradual dimming occurring clearly with the faint sources, while the brighter sources were not affected. This could be identified based on the abrupt corrections on the pointing, as the field of view was drifting due to an imprecise telescope tracking. Figure B.1 shows a sequence of eight exposures taken between 03:43:54 and 03:44:02 UT (i.e. after the occultation segment). Figure B.2 shows that the phantom target star smoothly disappears gradually. We successfully fit this gradual disappearance using the same instrumental response as in Sect. 4.3. The gradual ingress and egress during the occultation and the gradual disappearance of the phantom stars favour the hypothesis of an instrumental effect.

The recording Watec 910HX is an analogue video camera that is widely used in the network of occultation observers worldwide. This low-light video camera with integrating functions has an optional noise reduction filter called 3DNR (adjustable between 0 and 100%). This filter is activated at 50% level by default, and is automatically activated each time the camera settings are reset. The exact operation of this proprietary filter is unknown, but some tests and documentation have shown that it is a kind of running-average filter that smooths all the changes that occur at low light levels. For occultation work, this filter is turned off by the observers, and this setting is normally kept in memory for the next uses.

This gradual ingress and egress behaviour was recently observed during other stellar occultation events that were recorded by different observers using Watec 910HX cameras. In the light of these new data, we performed several tests trying to explain this behaviour. Figure B.3 illustrates the results of tests that successfully reproduced the desired behaviour, that is, the gradual dimming (over a few exposures) of the phantoms of the faint sources only.

To conclude, two reasons indicate that the observed instrumental effects are caused by the 3DNR filter, although the observer has no recollection nor knowledge of having it activated:

1. Cases of gradual occultation curves are regularly identified by IOTA coordinators to be the result of an unwanted activation of this filter. It appeared in particular that it is rather easy for the observer to reset the camera settings instead of exiting the camera menu, with the consequence of unintentionally reactivating the 3DNR filter.
2. Additional tests performed with a 910HX and a 3DNR filter at 50% level were able to reproduce the exact behaviour described in Fig. B.1. The same tests using a 3DNR filter set to off showed no anomaly.

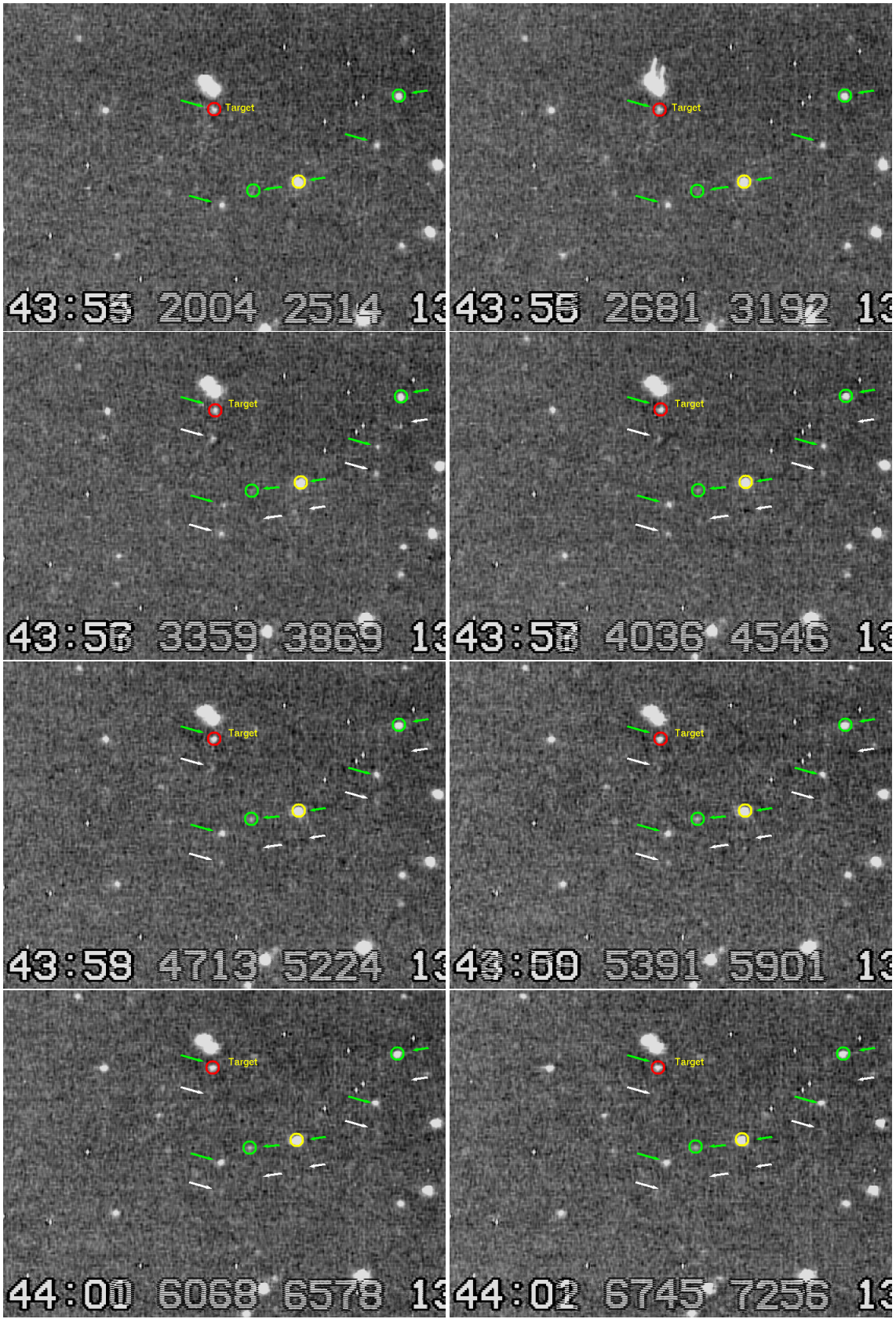

**Fig. B.1.** Sequence of eight sub-frame images taken a few minutes after the event (between 03:43:54 and 03:44:02 UT). In each frame, the target star is circled in red and the guide and calibration stars are circled in yellow and green, respectively. Because of the tracking correction (due to imprecise telescope tracking), the stars move abruptly on the CCD during the exposure at 03:44:55 (the sub-frame only shows the minutes and seconds). The green arrows point towards the real stellar positions. In the 03:43:57 exposure, we see phantoms of the dimmest stars (remnants from the previous exposures), which are shown by the white arrows. These phantoms are seen for at least five images, up to the 03:44:01 exposure.

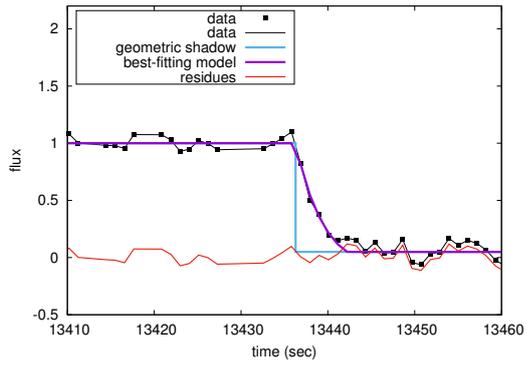

**Fig. B.2.** Gradual disappearance of the target phantom modelled assuming an (RC) electronic filter response with a time constant $\tau = 3.21$ s.

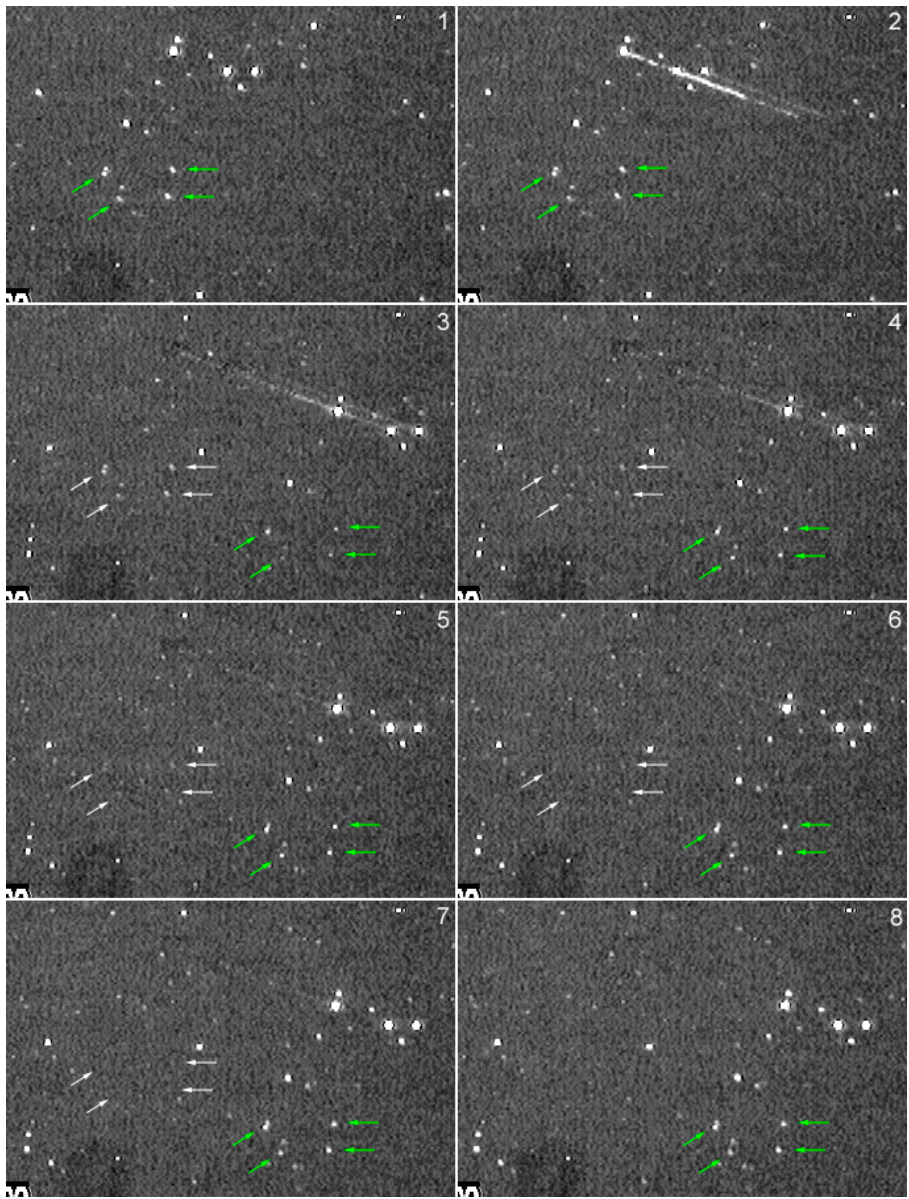

**Fig. B.3.** Eight successive integrations of 1.28 s with a Watec 910HX and 3DNR set to ON at 50% level, during an abrupt correction of the star field centring. The green arrows point towards several faint reference stars. The white arrows point towards ghost images of these faint stars at their first position, fading gradually with time. Bright stars are not affected by the filter.